\documentclass[webpdf,contemporary,large,namedate]{oup-authoring-template}%

\usepackage{lmodern}

\begin{document}
\journaltitle{TBD}
\DOI{TBD}
\copyrightyear{2024}
\pubyear{2024}
\access{Advance Access Publication Date: Day Month Year}
\appnotes{Preprint}
\firstpage{1}

\title[Pangene graphs]{Exploring gene content with pangene graphs}
\author[1,2,3,$\ast$]{Heng Li\ORCID{0000-0003-4874-2874}}
\author[2]{Maximillian Marin\ORCID{0000-0002-9108-3328}}
\author[2,4]{Maha Reda Farhat\ORCID{0000-0002-3871-5760}}
\address[1]{Dana-Farber Cancer Institute, 450 Brookline Ave, Boston, MA 02215, USA}
\address[2]{Harvard Medical School, 10 Shattuck St, Boston, MA 02215, USA}
\address[3]{Broad Insitute of Harvard and MIT, 415 Main St, Cambridge, MA 02142, USA}
\address[4]{Pulmonary and Critical Care Medicine, Massachusetts General Hospital, Boston, USA}
\corresp[$\ast$]{Corresponding author. \href{mailto:hli@ds.dfci.harvard.edu}{hli@ds.dfci.harvard.edu}}


\abstract{
\sffamily\footnotesize
\textbf{Motivation:}
The gene content regulates the biology of an organism.
It varies between species and between individuals of the same species.
Although tools have been developed to identify gene content changes in bacterial genomes,
none is applicable to collections of large eukaryotic genomes such as the human pangenome.\vspace{0.5em}\\
\textbf{Results:}
We developed pangene, a computational tool to identify gene orientation, gene order
and gene copy-number changes in a collection of genomes.
Pangene aligns a set of input protein sequences to the genomes,
resolves redundancies between protein sequences
and constructs a gene graph with each genome represented as a walk in the graph.
It additionally finds subgraphs, which we call bibubbles, that capture gene content changes.
Applied to the human pangenome, pangene identifies known gene-level variations
and reveals complex haplotypes that are not well studied before.
Pangene also works with high-quality bacterial pangenome and reports similar numbers of core and accessory genes
in comparison to existing tools.\vspace{0.5em}\\
\textbf{Availability and implementation:}
Source code at \url{https://github.com/lh3/pangene};
pre-built pangene graphs can be downloaded from \url{https://zenodo.org/records/8118576}
and visualized at \url{https://pangene.bioinweb.org}.
}

\maketitle

\section{Introduction}
A human genome contains about 20,000 protein-coding genes.
A small number of them have frequent copy-number or gene order changes in the human population~\citep{Sudmant:2010aa,Handsaker:2015ur}.
These genes are under fast evolution
and may be responsible for immune responses,
affecting brain functionality~\citep{Ju:2016aa} and drug metabolism~\citep{Taylor:2020aa},
or associated with known diseases~\citep{Mercuri:2022aa}.
They may have profound biological and biomedical implications.

Thanks to the recent advances in sequencing technologies \citep{Wenger_2019} and assembly algorithms~\citep{Nurk:2020we,Cheng:2021aa,Rautiainen:2023aa},
we can routinely achieve haplotype-resolved assembly over genes under copy-number or order changes.
We have also developed algorithms to construct pangenome sequence graphs that encode variations between genomes.
However, how to identify these gene-level variations is not straightforward.
Among the three whole-genome pangenome construction tools used by the Human Pangenome Reference Consortium (HPRC),
minigraph~\citep{Li:2020aa} and minigraph-cactus~\citep{Hickey:2023aa} are
unable to align through complex genomic regions and may miss genes in long segmental duplications;
PGGB~\citep{Garrison2023.04.05.535718} collapses paralogous genes which makes it difficult to study individual paralogs.
In addition, all three tools do not directly reveal how genomic variations affect genes.
To study gene-level variations, HPRC
had to manually annotate genes on each haplotype~\citep{Liao:2023aa} which is a time-consuming process.
PGR-TK~\citep{Chin:2023aa} reconstructs local haplotype structures from genomic sequences
but it is does not directly model genes and is not intended for whole-genome data.
The current human pangenome tooling is not designed for studying gene variations.

In contrast, research on bacterial pangenome
focuses on protein-coding genes instead of genome sequences, to the point that
in the literature, a bacterial ``pangenome'' often refers to the collection of protein-coding genes.
Several high-quality tools have been developed for constructing the gene content of bacterial genomes~\citep{Page:2015aa,Ding:2018aa,Tonkin-Hill:2020aa,Gautreau:2020aa,Zhou:2020aa}.
In a nutshell, they start with \emph{ab initio} gene annotation in each genome,
cluster the resulted protein sequences,
and then post-process clusters to identify orthologous genes
and to fix issues caused by imperfect assembly, annotation or clustering~\citep{Tonkin-Hill:2023aa}.
These bacterial pangenome tools however have not considered splicing,
multiple isoforms, frequent segmental duplications and the much larger size of the human genome.
They have not been shown to work with human pangenome data.

Here we developed pangene, a new computational tool to explore the gene content of a pangenome.
Unlike bacterial pangenome pipelines, pangene effectively annotates protein-coding genes
by aligning protein sequences to each genome with miniprot~\citep{Li:2023ac}.
As miniprot can align through in-frame stop codons and frameshifts,
this procedure simplifies gene annotation and is robust to insertion/deletion errors in the input genomes.
Furthermore, pangene constructs a bidirected gene graph and can capture inversions
missed by bacterial pangenome tools.
It also provides an algorithm to
identify gene copy-number or gene order variations.
Pangene is optimized for human genomes and also works for bacterial genomes.

\section{Methods}

\begin{table}[!tb]
\caption{Notations\label{tab:sym}}
\begin{tabular*}{\columnwidth}{@{\extracolsep\fill}ll@{\extracolsep\fill}}
\toprule
Notation & Description \\
\midrule
$V$ & Set of genes \\
$v,w,u$ & Genes; vertices in a bidirected graph \\
$X$ & Set of oriented genes; $X=V\times({\rm >},{\rm <})$ \\
$x,y,z$ & Oriented genes; vertices in a directed graph \\
${\rm >}v,{\rm <}v$ & Oriented genes \\
$\overline{x}$ & Reverse complement of $x$; $\overline{{\rm >}v}={\rm <}v,\overline{{\rm <}v}={\rm >}v$ \\
$\nu(x)$ & The gene behind $x$; $\nu({\rm >}v)=\nu({\rm <}v)=v$ \\
\botrule
\end{tabular*}
\end{table}

Pangene takes a set of protein sequences and multiple genome assemblies as input,
and outputs a graph in the GFA format~\citep{Li:2020aa}.
It involves two steps: aligning the set of protein sequences to each input assembly with miniprot~\citep{Li:2023ac},
and deriving a graph from the alignment with each contig encoded as a walk of genes.
Pangene provides utilities to classify genes into core genes that are present in most of the input genomes, or accessory genes otherwise.
Pangene can also identify generalized ``bubbles'' in the graph, which represent local gene order,
gene copy-number or gene orientation variations among the input genomes.

\begin{figure}[b!]
\centering
\includegraphics[width=.48\textwidth]{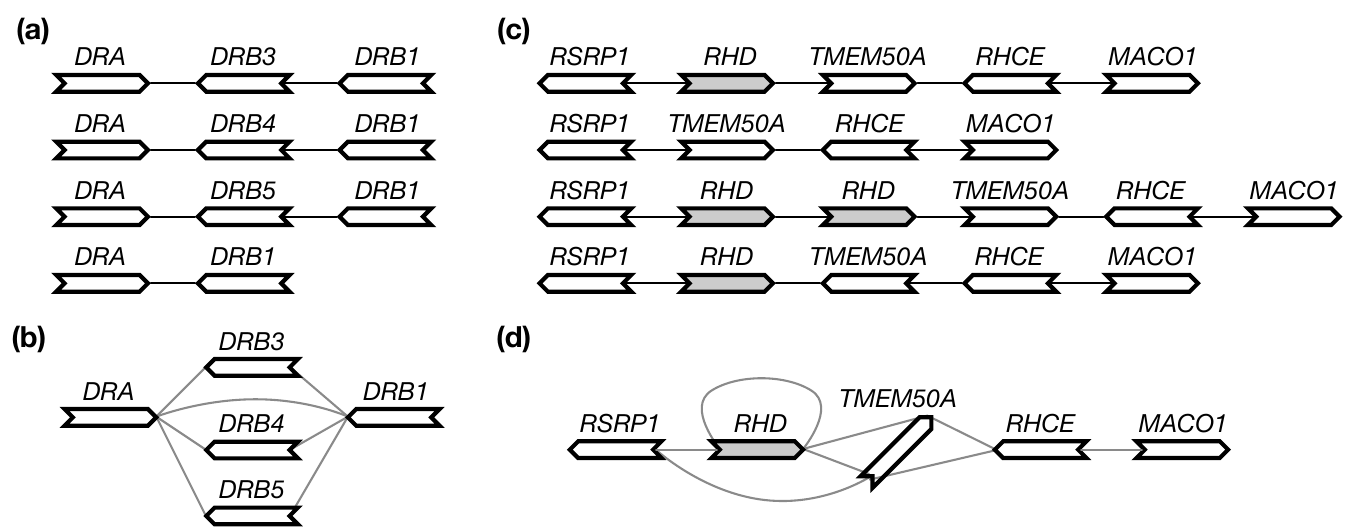}
\caption{Examples of pangene graphs. {\bf (a)} Human haplotypes around the
\emph{HLA-DRB1} gene. {\bf (b)} The pangene graph around \emph{HLA-DRB1}. {\bf
(c)} Human haplotypes around the \emph{RHD} gene. \emph{RHD} has copy-number
changes and \emph{TMEM50A} may be inverted. {\bf (d)} The corresponding pangene
graph.}\label{fig:ex1}
\end{figure}

Given perfect gene annotation and orthology assignment between the genes, the pangene graph construction algorithm is conceptually simple:
it takes an orthologous group as a vertex and adds an edge between two groups
if on an input genome, a gene in one group is adjacent to a gene in the other group (Fig.~\ref{fig:ex1}).
The practical difficulty is to obtain accurate annotation in the presence of redundant sequences, paralogous genes and errors in assembly or alignment.

An important use of pangenome graphs is to identify genomic variations represented as ``bubbles''.
While bubbles have been extensively studied in directed graphs~\citep{DBLP:conf/wabi/OnoderaSS13},
they have not been rigorously defined on bidirected graphs graphs that are the necessary outputs of assemblers and pangenome constructors.
If the field is to move towards the more generalized bidirected graph as a representation of genomic variation developing the concept of bubbles in this context will be necessary.
Hence, how to define and to find ``bubbles'' in bidirected graphs will be a major topic of this article.

\subsection{Defining pangene graphs}

\begin{figure}[t!]
\centering
\includegraphics[width=.48\textwidth]{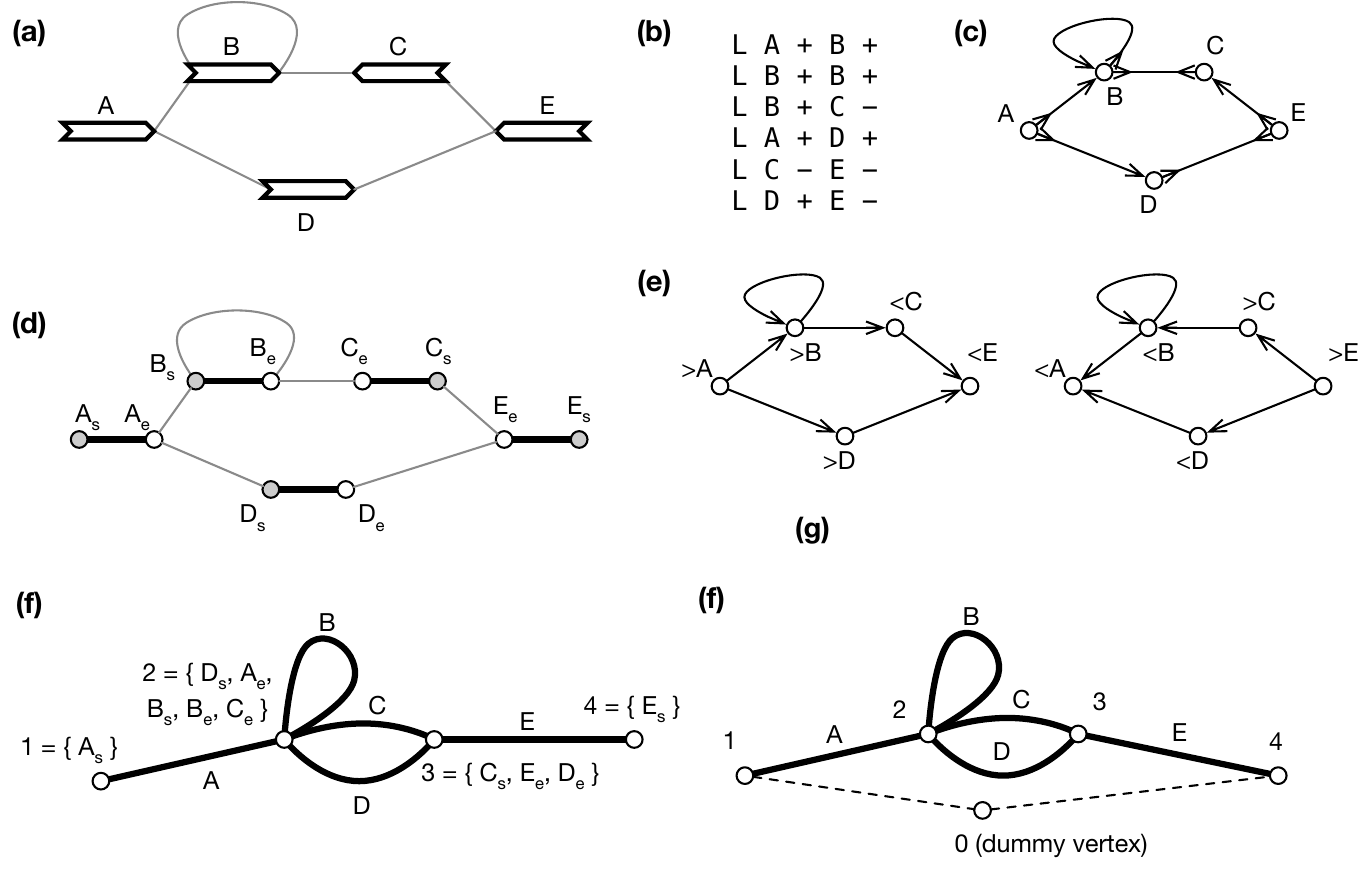}
\caption{An example of a gene graph. {\bf (a)} A gene graph for visualization.
{\bf (b)} The corresponding GFA format.
{\bf (c)} Bidirected graph.
{\bf (d)} Biedged graph.
{\bf (e)} Directed graph.
{\bf (f)} Net graph, which loses information.
{\bf (g)} A depth-first traversal of the net graph with ``0'' representing a super node
not in the original net graph.
(a) through (e) are equivalent representations.}\label{fig:exa1}
\end{figure}

Let $V$ be the set of genes
and $X=V\times\{{\rm >},{\rm <}\}$ be the set of oriented genes.
If $v\in V$ is a gene, $x=(v,{\rm >})\in X$, or simply $x={\rm >}v$, denotes an oriented gene
and $\nu(x)=v$ gives the underlying gene behind oriented gene $x$.
$\overline{x}$ is the reverse complement of $x$.
i.e., if $x={\rm >}v$, $\overline{x}={\rm <}v$, and vice versa.
Throughout this article, we will use symbol $u$, $v$ or $w$ to denote a gene
and use $x$, $y$ or $z$ to denote an oriented gene (Table~\ref{tab:sym}).

Let $T$ be the set of input contigs.
They harbor oriented genes in $X$.
2-tuple $(x,y)\in X\times X$ is said to be \emph{supported} by contig $t\in T$
if $y$ immediately follows $x$ on $t$ or, due to the strand symmetry of DNA, $\overline{x}$ immediately follows $\overline{y}$.

A pangene graph is usually visualized as Fig.~\ref{fig:exa1}a~\citep{Wick:2015qf}
and can be mathematically defined in a few equivalent ways.
It can be formulated as a directed graph $G_D=(X,E)$ where $E\subset X\times X$ is the set of edges.
In the context of pangene graphs, $(x,y)$ is an edge, also written as $x\to y$, if $(x,y)$ is supported by a contig.
Notably, if $x\to y$ is an edge, $\overline{y}\to\overline{x}$ must also be an edge.
In graph theory, this property is called the \emph{skew symmetry}.
Fig.~\ref{fig:exa1}e shows an example of a directed pangene graph
which can be described in the GFA format (Fig.~\ref{fig:exa1}b).
In $G_D$, it is possible that $y$ can reach both $x$ and $\overline{x}$.
When this happens, $x$ or $\nu(x)$ is said to be an \emph{inversion}.

A pangene graph can also be thought as a bidirected graph $G_B=(V,E)$ where the definition of $E$ is the same as above.
An edge in $E$ is effectively associated with two directions.
For example, $({\rm >}v,{\rm <}w)\in E$, also written as $v{\rm ><}w$,
suggests the edge between $v$ and $w$ has a first direction ``${\rm >}$'' towards vertex $v$ and a second direction ``${\rm <}$'' towards vertex $w$.
Fig.~\ref{fig:exa1}c shows the bidirected equivalence of Fig.~\ref{fig:exa1}a.

In a directed graph, $x\to y\gets z$ is not a path because $x$ cannot reach $z$.
The similar constraint is applied to the bidirected graph $G_B$.
For example, $v{\rm >>}u{\rm <<}w$ is not a path because both $v$ and $w$ go into $u$.
$v{\rm >>}u{\rm ><}w$ is a path, which can also be expressed as a path in $G_D$: $({\rm >}v,{\rm >}u,{\rm <}w)$.


As pangene constructs the graph from protein-to-genome alignment~\citep{Li:2023ac},
it also attaches alignment scores to edges.
Let $T_{x\to y}\subset T$ be the set of contigs that support $x\to y$
and $S(x|y,t)$, $t\in T_{x\to y}$, be the alignment score of $x$ when $x\to y$ is supported by contig $t$.
$S(x|y)$ is calculated as
$$
S(x|y)=\frac{1}{|T_{x\to y}|}\sum_{t\in T_{x\to y}}S(x|y,t)
$$
which is the average score of $x$ among contigs supporting $x\to y$.

\subsection{Obtaining the protein set}

Pangene requires a set of protein sequences as input which may have redundancies.
In principle, if every input genome is annotated, we may merge all annotated protein sequences
and align to all genomes with miniprot.
This strategy works when there are a small number of diverged genomes.
If there are many similar genomes, all with annotations,
a faster way is to cluster protein sequences with CD-HIT~\citep{Li:2006aa} or MMseqs2~\citep{Schneider:2017aa},
and take a representative protein sequence from each cluster to generate the protein set.

Most human genomes do not have high-quality gene annotation.
We may generate the protein set from the reference gene annotation supplemented
with protein-coding genes or diverged alleles not present in the reference genome, such as \emph{HLA-DRB3}.
Pangene allows one gene to be associated with multiple protein sequences.
If protein names follow format {\tt <GeneID>:<ProteinID>},
pangene will select one protein for each {\tt <GeneID>} and use {\tt <GeneID>} as the segment name in the output GFA file.

\subsection{Selecting non-orthologous genes}

The central problem in graph construction is how to consistently annotate multiple genomes with a subset of input proteins.
A naive way to annotate a genome is to align all protein sequences to the genome
and if there are multiple proteins mapped to the same locus, select the best scoring protein as the annotation.
This simple strategy does not work if there are orthologous proteins in the input protein set.
For example, suppose gene $v$ and $w$ are orthologous to each other and are thus aligned to the same locus in each genome.
When some genomes match $v$ better and the rest match $w$ better,
both $v$ and $w$ will be considered as accessory genes that are not present in all genomes.
However, the preferred choice is to elect one of them as a core gene.

Pangene selects non-orthologous genes as follows.
For gene $v$, pangene inspects its best hit in each genome (hashes of gene names are used to break ties)
and counts $b(v)$ the number of genomes where $v$ is aligned better than all other genes overlapping with $v$.
Pangene then traverses each gene $v$ in the descending order of current $b(v)$.
If $b(v)>0$, select $v$ as a non-orthologous gene and for each genome where the best mapping of $w$ overlaps with $v$ but is aligned better than $v$, reduce $b(w)$ by one.
This way, if $v$ and $w$ are orthologs, their best mapping positions likely overlap and thus only one of them will be selected;
if $v$ and $w$ are paralogs, their best mapping positions usually differ and thus both will be selected.

\begin{figure}[tb!]
\centering
\includegraphics[width=.48\textwidth]{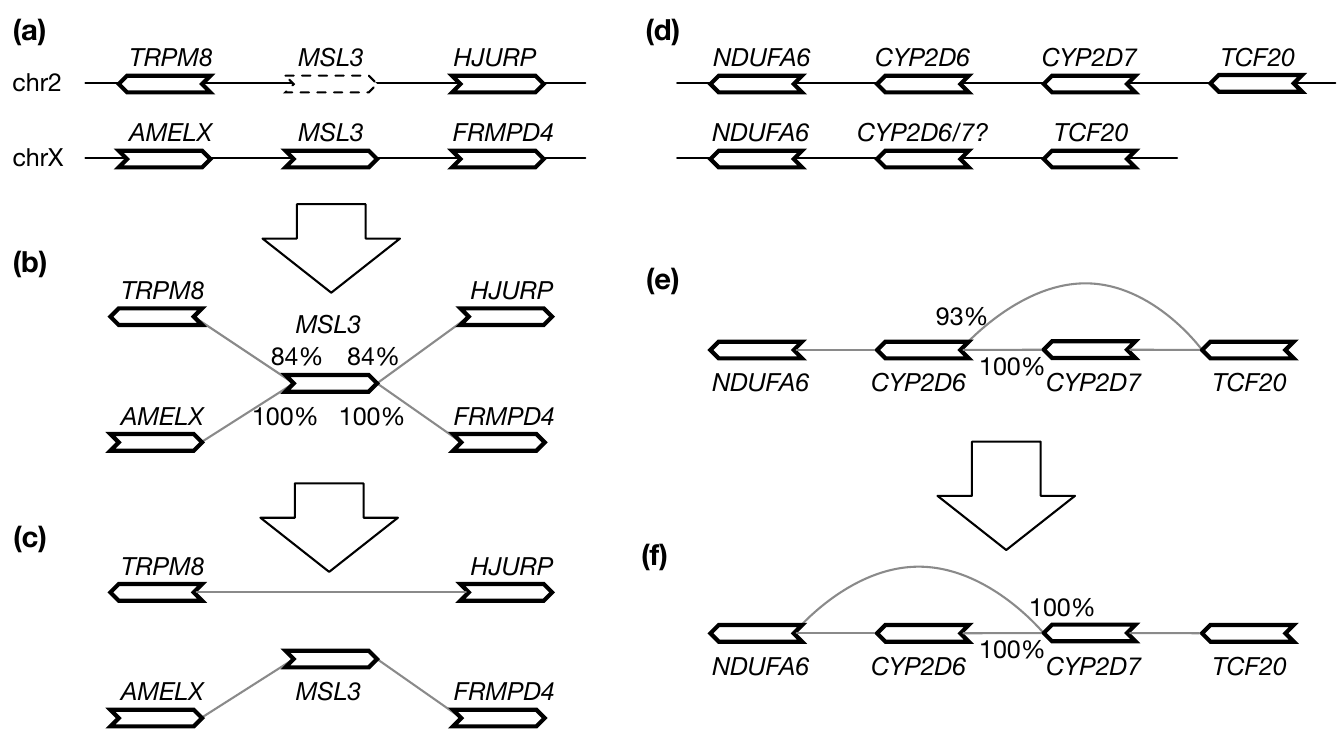}
\caption{Graph-based annotation adjustment.
{\bf (a)} \emph{MSL3} is located on chromosome X but has a weaker hit on chromosome 2 when the gene
is aligned to a male haplotype without chromosome X.
{\bf (b)} The initial pangene graph. Percentages indicate alignment identity.
{\bf (c)} The adjusted graph after filtering out the \emph{MSL3} hit on chr2.
{\bf (d)} Two haplotypes involving \emph{CYP2D6} and \emph{CYP2D7} which are paralogous genes.
The second haplotype should have \emph{CYP2D7} only but because \emph{CYP2D6} is longer, the \emph{CYP2D6} alignment score is higher than the \emph{CYP2D7} score.
{\bf (e)} The initial pangene graph assuming the second haplotype has \emph{CYP2D6}.
{\bf (f)} The adjusted graph.
}\label{fig:resolve}
\end{figure}

\subsection{Adjusting pangene graphs}\label{sec:adj}

Pangene constructs an initial graph using genes selected from the last section.
It applies additional heuristics to improve gene annotations based on the graph topology and alignment scoring
in two special cases.

First, a gene may be wrongly mapped to its paralogs if its best position is missing.
Suppose a gene on chromosome X has a second best hit on an autosome.
If the alignment score of the autosomal hit is lower than the hit to chromosome X by 3\%
miniprot will only report the hit on chromosome X.
However, because a phased male human assembly does not have chromosome X,
the gene will be mapped to the autosome (Fig.~\ref{fig:resolve}a).
This false hit will lead to spurious edges in the initial graph (Fig.~\ref{fig:resolve}b).
To solve this problem, pangene marks $x\to y$ as a false edge
if there exists $x\to z$ such that $y$ and $z$ are on different contigs in every input genome and $S(x|y)<S(x|z)\cdot r_1$ ($r_1$ defaults to 95\%).
Pangene then filters out all alignment of $x$ that are incident to a false edge $x\to y$ (Fig.~\ref{fig:resolve}c).

Second, similar to the first case, if the best position of a gene is missing due to a real deletion in evolution,
this gene may interfere with the alignment of paralogous genes (Fig.~\ref{fig:resolve}d)
and leads to a graph (Fig.~\ref{fig:resolve}e) inconsistent with manuscript annotation~\citep{Liao:2023aa}.
To improve this case, pangene marks $x\to y$ to be low priority
if there exists $x\to z$ such that $y$ and $z$ are on the same contig in an input genome and $S(x|y)<S(x|z)\cdot r_2$ ($r_2$ defaults to 98\%).
Pangene then marks all $x$ alignment that are incident to a low-priority edge $x\to y$ to have low priority as well
and prefers a gene without the low-priority mark.
In Fig.~\ref{fig:resolve}e, $x={\rm <}$\emph{CYP2D6}, $y={\rm <}$\emph{TCF20} and $z={\rm <}$\emph{CYP2D7}.
${\rm <}$\emph{CYP2D6} is marked to have low priority on the second contig.
This results in a graph in Fig.~\ref{fig:resolve}f that is more consistent with the known evolution at this locus~\citep{Liao:2023aa}.

Both cases may also be caused by processed pseudogenes.
When a protein-coding gene has a recent pseudogene that has a near identical sequence,
the two heuristics may not work due to the score ratio threshold.
By default, pangene filters out an alignment without splicing as a potential pseudogene alignment if the protein has a spliced alignment in other samples.
Pangene also has an option to drop a protein if it does not have spliced alignment in any sample.
For human samples, these strategies can reduce false connections between chromosomes.

\subsection{Past work on bubble finding}

The concept of bubble is perhaps first coined by \citet{Zerbino:2008uq} in the context of assembly graphs.
A bubble was initially meant to be simple in that paths through a bubble do not share vertices except the start and the end vertices.
\citet{DBLP:conf/wabi/OnoderaSS13} introduced superbubble to allow more complex acyclic topology
and found a quadratic algorithm to identify all superbubbles.
The time complexity was improved to $O(|E|\log|E|)$ by \citet{Sung:2015aa}
and then to linear by \citet{DBLP:journals/tcs/BrankovicIKMPV16}.

A subgraph induced by a superbubble is acyclic, but in a pangene graph or a variation graph,
we also care about cyclic subgraph.
\citet{Gartner:2018aa} defined weak superbubble without the acyclic condition,
and presented a linear algorithm to identify all weak superbubbles~\citep{DBLP:journals/algorithms/GartnerS19}.
Interestingly, more than 20 years before this work,
\citet{DBLP:conf/pldi/JohnsonPP94} had already come up with a linear algorithm to effectively find weak superbubbles in the context of compiler design,
which will be the basis of our algorithm.

Weak superbubble is defined on directed graphs, not on bidirected graphs.
If we apply the previous algorithms to directed graph $G_D$,
we will find most superbubbles twice on opposite strands.
More importantly, superbubble cannot capture the visual intuition in the presence of inversions.
For example, in Fig.~\ref{fig:inv}a, {\tt >A} and {\tt <C} enclose a subgraph that looks like a bubble,
but the equivalent directed graph (Fig.~\ref{fig:inv}f) does not have a superbubble when {\tt >A} and {\tt <A} are taken as different vertices.
\citet{Dabbaghie:2022aa} implemented the algorithm by \citet{DBLP:conf/wabi/OnoderaSS13} for sequence graphs,
but how the issues above are handled is not apparent.
The closest concept to weak superbubble in bidirected graphs is snarl~\citep{Paten:2018aa},
which leads to a bidirected subgraph that can be separated from the rest of the graph.
Snarl does not require reachability which is necessary in defining superbubbles.
To this end, there are no equivalent definition of weak superbubble in directed graphs.

\subsection{Defining generalized bibubbles}


The definition of ``bubble'' here tries to capture the visual intuition.
It is inspired by \citet{DBLP:conf/wabi/OnoderaSS13}.
Let $U(x,y)\subset V$ be the set of vertices that are reachable from $x$ without passing through $x$, $\overline{x}$ or $y$.
A \emph{generalized bidirected bubble}, or a \emph{generalized bibubble} in short, is a pair $(x,y)\in X\times X$ such that
i) $U(x,y)=U(\overline{y},\overline{x})\not=\emptyset$;
ii) $\forall v\in U(x,y)$, there is a walk from $x$ to $y$ that passes $v$;
iii) there does not exist $z\in U(x,y)\times\{{\rm >},{\rm <}\}$
that satisfies $U(x,z)=U(\overline{z},\overline{x})$ or $U(z,y)=U(\overline{y},\overline{z})$.
Due to the skew symmetry, if $(x,y)$ is a generalized bibubble, $(\overline{y},\overline{x})$ is also a generalized bibubble.

\begin{figure}[t!]
\centering
\includegraphics[width=.48\textwidth]{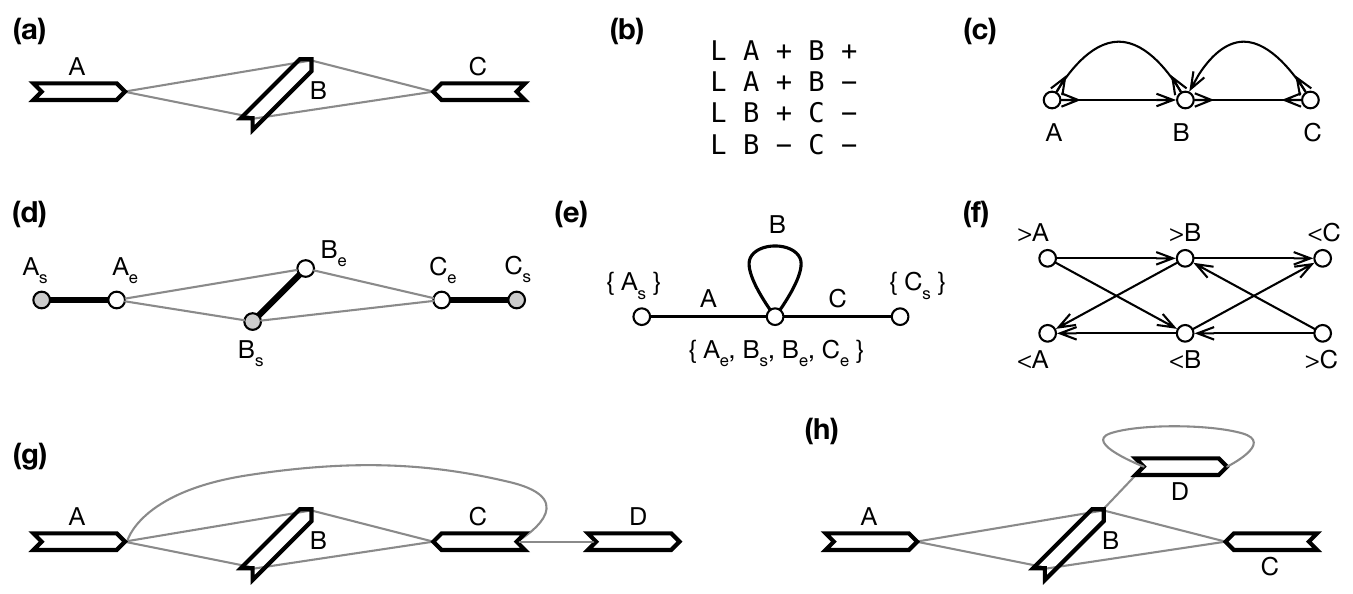}
\caption{Example of gene graphs with inversions.
{\bf (a)} A gene graph for visualization. ({\tt >A},{\tt <C}) is a generalized bibubble.
{\bf (b)} The corresponding GFA format.
{\bf (c)} Bidirected graph.
{\bf (d)} Biedged graph.
{\bf (e)} Net graph, which loses information.
{\bf (f)} Directed graph.
{\bf (g)} A second gene graph with ({\tt >A},{\tt >D}) being a generalized bibubble.
({\tt >A},{\tt <C}) is not a generalized bibubble because $U$({\tt >A},{\tt <C})=$\{${\tt B},{\tt C}$\}$
but $U$({\tt >C},{\tt <A})=$\{${\tt B}$\}$.
{\bf (h)} A third gene graph without any generalized bibubbles.
}\label{fig:inv}
\end{figure}

In the definition of $U(x,y)$, we intentionally allow walks passing through $\overline{y}$;
otherwise $U$({\tt >A},{\tt <C}) in Fig.~\ref{fig:inv}g would become $\{${\tt B}$\}$, making ({\tt >A},{\tt <C}) a generalized bibubble.
This example also suggests $(x,y)$ would not be a generalized bibubble if $x\in U(x,y)$ or $y\in U(x,y)$.
Condition ii) above aims to exclude examples such as Fig.~\ref{fig:inv}h.
Condition iii) requires a bibubble to be minimal.
As a result, it is not possible for $(x,y)$ and $(x,z)$ both to be generalized bibubbles unless $y=z$.
Generalized bibubbles are also nested in that
if $(x,x')$ and $(y,y')$ are generalized bibubbles and $U(x,x')\cap U(y,y')\not=\emptyset$,
then $U(x,x')\subset U(y,y')$ or $U(y,y')\subset U(x,x')$.

For a generalized bibubble $(x,y)$, let $\overline{U}(x,y)=V\setminus U(x,y)\setminus\{\nu(x),\nu(y)\}$.
No vertex in $\overline{U}(x,y)$ can reach vertices $U(x,y)$
because otherwise $w\in\overline{U}(x,y)$ reachable from $x$ (or from $\overline{y}$)
would not pass through $y$ (or $\overline{x}$) and would be added to $U(x,y)$ (or $U(\overline{y},\overline{x})$)
-- this would violate the definition of $U(x,y)$.
This observation indicates that removing $\nu(x)$ and $\nu(y)$ would separate $U(x,y)$ from the rest of the graph.

A naive algorithm to identify all generalized bibubbles is to enumerate each pair of oriented genes
and then test whether the pair forms a generalized bibubble based on the definition.
This is an $O(|V|^2\cdot(|V|+|E|))$ algorithm, impractical for large graphs.
To describe a new algorithm for bubble finding, we will first introduce net graphs and cycle equivalence.

\subsection{Biedged graphs and net graphs}

Let $Z=V\times\{{\rm s},{\rm e}\}$, where symbol ``${\rm s}$'' represents the start of a gene
and symbol ``${\rm e}$'' represents the end.
Given a bidirected graph $G_B=(V,E)$, a \emph{biedged graph} $G_E=(Z,E_g,E_l)$ is constructed
with $E_g=\big\{\{(v,{\rm s}),(v,{\rm e})\}|v\in V\big\}$ and $E_l$ being the set of undirected connections between starts or ends of genes in $Z$~\citep{Paten:2018aa}.
Fig.~\ref{fig:exa1}d shows an example.
Note that a valid walk in a biedged graph cannot pass two consecutive edges in $E_l$.
Due to this restriction, many classical algorithms in graph theory are not applicable to a biedged graph.

A \emph{net graph} $G_N=(P,E_n)$ is constructed by contracting all edges in $E_l$.
As a result, a vertex in $P$ is a component connected by edges in $E_l$
and an edge represents a gene in $V$ (Fig.~\ref{fig:exa1}e and Fig.~\ref{fig:inv}e).
Because $E_n$ and $V$ has one-to-one relationship, we will also use $v$ to denote an edge in $G_N$.
A net graph is an undirected graph and is not equivalent to the bidirected graph $G_B$.
For example in Fig.~\ref{fig:exa1}a, there is not a walk that goes through both gene {\tt C} and {\tt D},
but in Fig.~\ref{fig:exa1}f, there is a walk involving both genes.

In graph theory, a set of edges in $G_N$ is a \emph{cut-set} if cutting through these edges
partitions $G_N$ into two disconnected subgraphs.
A key observation is that if $(x,y)$ is a bibubble in bidirected graph $G_B$,
$\{\nu(x),\nu(y)\}$ is a cut-set in $G_N$.

\subsection{Cycle equivalence}

Two edges $e_1$ and $e_2$ in $G_N$ are said to be \emph{cycle equivalent}
if every cycle containing $e_1$ contains $e_2$ and vice versa.
Cycle equivalent edges form equivalent classes, which can be determined in linear time~\citep{DBLP:conf/pldi/JohnsonPP94}.

For any pair of $e_1$ and $e_2$ that are cycle equivalent, $\{e_1,e_2\}$ is a cut-set.
To see this, let $p_{11}$ and $p_{12}$ are the two vertices incident to $e_1$.
Without losing generality, suppose $G_n$ is a connected graph.
If $G_n$ remained connected after removing $e_1$ and $e_2$,
there would exist a vertex $q$ that can reach both $p_{11}$ and $p_{12}$ without passing through $e_2$ (because $e_2$ has been removed).
Then $e_1$ would be in a cycle with $q$ in the original $G_N$ but without passing through $e_2$,
violating the cycle equivalence of $e_1$ and $e_2$.

Conversely, if $\{e_1,e_2\}$ is a cut-set, $e_1$ and $e_2$ are cycle equivalent.
To prove this, let $\{e_1,e_2\}$ partition the vertex set $P$ into $P_1$ and $P_2$.
If $e_1$ and $e_2$ were not cycle equivalent, there must exist a cycle that only contains $e_1$ but not $e_2$.
Then cutting both $e_1$ and $e_2$ would still keep $P_1$ and $P_2$ connected due to this cycle,
violating that $\{e_1,e_2\}$ is a cut-set.

\begin{figure}[t!]
\centering
\includegraphics[width=.48\textwidth]{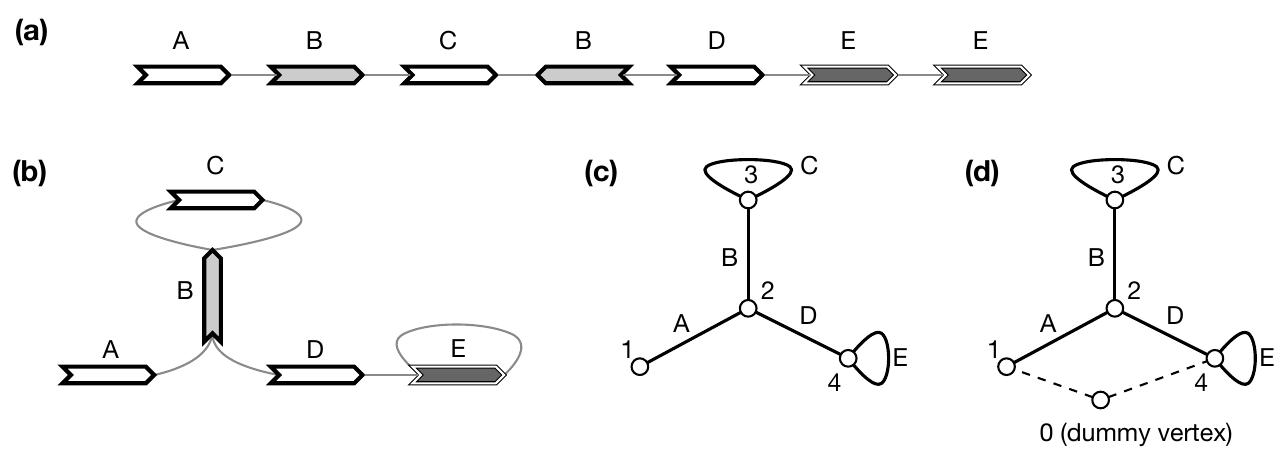}
\caption{Example of gene graphs that complicate net graph based bubble finding.
{\bf (a)} Gene annotation on a contig.
{\bf (b)} The corresponding pangene graph.
{\bf (c)} Net graph with the dummy vertex ``0'' connected to ``1'' only.
{\bf (d)} Net graph with the dummy vertex connected to genes at the ends of the contig.
}\label{fig:complex}
\end{figure}

To this end, $\{e_1,e_2\}$ is a cut-set if and only if $e_1$ and $e_2$ are cycle equivalent.
Then if $(x,y)$ is a generalized bibubble in $G_B$,
$\nu(x)$ and $\nu(y)$ are cycle equivalent in $G_N$.
The reverse is not always true.
For example, in Fig.~\ref{fig:exa1}f, ``{\tt C}'' and ``{\tt D}'' are cycle equivalent,
but ({\tt >C},{\tt >D}) or in any orientation is not a generalized bibubble.
Similarly, ``{\tt A}'' and ``{\tt B}'' are cycle equivalent in Fig.~\ref{fig:complex}c but they do not enclose generalized bibubbles.


\subsection{Identifying generalized bibubbles}

Given a bidirected graph $G_B$, pangene transforms it to a net graph $G_N$
and assigns a class number to each gene $v$ such that $v$ and $w$ are in the same class if they are cycle equivalent in $G_N$.
For each oriented gene $x$ that has multiple edges,
pangene applies a breadth-first search (BFS) to traverse up to $m$ genes (100 by default) reachable from $x$.
During BFS, pangene collects $y$ in the same equivalent class and does not traverse beyond $y$.
It then checks each $(x,y)$ using the definition of generalized bibubble.
Because $x$ can only appear in one generalized bibubble, pangene stops testing other $(x,\cdot)$ pairs
if it has found a generalized bibubble.

The time complexity in the worst case is $O(|V|+|E|+b\cdot m^2)$, where $b$ is the number of genes with multiple incoming or outgoing edges
and $m$ is effectively the maximum bibubble size.
On real pangene graphs, the time complexity is close to $O(|V|+|E|+b\cdot m)$.
It takes pangene less than one second to find $\sim$200 generalized bibubbles
in a human graph including 20 thousand genes.

In addition to cycle equivalence, \citet{DBLP:conf/pldi/JohnsonPP94}
described another linear algorithm to find single-entry-single-exist (SESE) regions in $G_N$
provided that $G_N$ contains a vertex $p_0$ that puts every other vertex in $G_N$ in a cycle with $p_0$.
A SESE region is intuitively a bubble-like subgraph in $G_N$.
Pangene also implemented this algorithm.
It introduced $p_0$ that connects all vertices with degree one in $G_N$
and connects the first and the last gene on each reference chromosome (Fig.~\ref{fig:exa1}f and Fig.~\ref{fig:complex}d).
Because a SESE region does not always correspond to a generalized bibubble in $G_B$,
pangene still has to test each SESE region using the definition of generalized bibubble.
The time complexity of this second algorithm is $O(|V|+|E|+b'm^2)$ where $b'$ is the number of
SESE regions in $G_N$.
It is faster in theory,
but pangene still uses the first algorithm by default
as the placement of $p_0$ is heuristic which caused one false negative in a human pangene graph.

\begin{figure*}[b!]
\centering
\includegraphics[width=\textwidth]{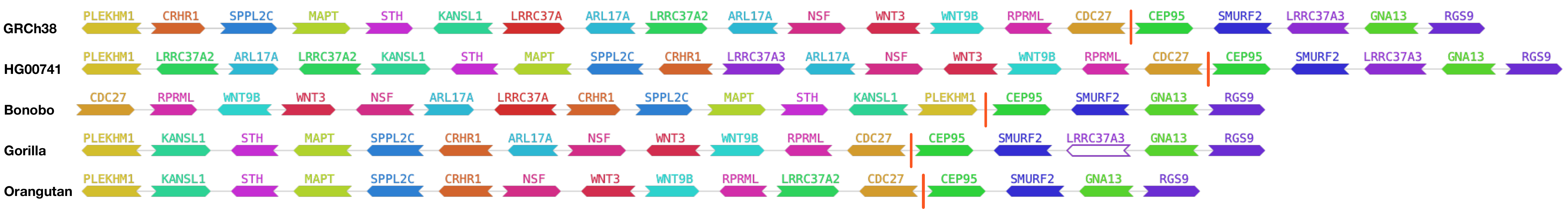}
\caption{Human and great ape haplotypes around \emph{LRRC37A*} genes.
A red vertical bar indicates the two genes on each side are not adjacent on the genome.
In the plot, each genome contains two blocks of genes.
The genomic distance between the two blocks is 17.3 Mb in human, 61.3 Mb in bonobo,
17.6 Mb in gorilla or 2.8 Mb in orangutan.
GRCh38 has \emph{LRRC37A} and \emph{LRRC37A2} in 17q21.31 and \emph{LRRC37A3} in 17q24.1.
HG00741 has a known inversion around \emph{MAPT}.
All great apes have only one \emph{LRRC37A*} gene.
Chimpanzee and bonobo have the same gene haplotype.
The two orangutan subspecies also have the same haplotype.
The \emph{LRRC37A3} alignment in gorilla has two in-frame stop codons.
It is challenging to distinguish \emph{LRRC37A*} paralogs
as they are similar in sequence and lack synteny with nearby genes.
}\label{fig:17inv}
\end{figure*}

\section{Results}

We applied pangene to HPRC samples and manually inspected subgraphs around \emph{CYP2D6}, \emph{RHD}, \emph{C4}, \emph{HLA-DRB1} and KIR genes
which have been curated for these samples~\citep{Liao:2023aa,Zhou2024.01.20.576452}.
We also looked at many other genes that are known to have gene-level variations~\citep{Sudmant:2010aa,Boettger:2012aa,Handsaker:2015ur}.
While these genes have not been manually annotated for HPRC samples,
we could check whether pangene captured known events such as
the \emph{MAPT} inversion and the \emph{ORM1} copy-number changes.

Due to the lack of genome-wide ground truth in human data,
we could only manually evaluate pangene in individual subgraphs.
To get a better sense of overall statistics,
we additionally applied pangene to two bacterial datasets
and compared the results to existing bacterial pangenome tools.

\subsection{Structural variants between two human genomes}

We downloaded GenCode comprehensive human annotation v45,
retained the protein-coding transcripts tagged with ``Ensembl\_canonical'' and filtered out readthrough transcripts (a transcript that joins two adjacent genes) and mitochondrial genes.
Only one transcript was selected per gene.
Because GenCode does not include HLA-DRB3, HLA-DRB4 and several KIR genes,
we manually added 20 {\it HLA-DRB} genes and alleles from the IPD-IMGT/HLA database and
15 KIR genes from the IPD-KIR database.
In the end, we constructed a protein set with 19,421 sequences.

We aligned the proteins to the human reference genome GRCh38~\citep{Schneider:2017aa} and T2T-CHM13~\citep{Nurk:2022up}
with miniprot~\citep{Li:2023ac} v0.12 under option ``{\tt --outs=0.97 -Iu}''
and constructed a pangene graph.
The resulting graph contains 19,088 genes and 19,390 edges.
We identified 91 generalized bubbles in the graph that contain 100 genes or fewer.
The bibubbles involved 691 genes affected by gene copy, gene order or gene orientation changes between the two genome.
These genes included many known events such as {\it PDPR}, {\it SMN2}, {\it CTAGE9}, {\it HPR}, {\it ORM1}, {\it CCL4}, {\it NCF1} and Amylase~\citep{Handsaker:2015ur,Sudmant:2010aa}.
Some of the large gene clusters affected by structural changes, such as {\it SMN2} and Amylase,
would not be easily captured by whole-genome alignment as they are not represented by colinear genome alignments.

\subsection{Analyzing 100 human haplotypes}

We obtained the assembly of CN1~\citep{Yang:2023aa}, YAO~\citep{He:2023aa} and 47 samples released by the Human Reference Pangenome Consortium~\citep{Liao:2023aa}.
Together with GRCh38 and T2T-CHM13, this gave us 100 human haplotypes.
We aligned the same set of proteins to the 100 haplotypes.
Pangene took less than one minute to construct a graph
and less than one second to identify 266 generalized bibubbles in the graph.
Many of the bibubbles were supported by one contig only.
Pruning edges supported by one contig resulted in a graph containing 209 generalized bibubbles.
Further dropping about 2,000 single-coding-exon genes led to a graph containing 163 bibubbles.
These included manually confirmed gene-level variations around {\it CYP2D6}, {\it C4} and
{\it RHD}~\citep{Liao:2023aa} and around {\it HLA-DBR1}~\citep{Zhou2024.01.20.576452}.

Pangene effectively annotates each input genome.
For a sanity check, we compared the pangene gene annotation to the GENCODE annotation.
Out of 19,043 protein coding genes in both annotations,
only four gene locations are different.
In one example, pangene assigned \emph{NUDT4B} to \emph{NUDT4}.
Pangene misses \emph{NUDT4B} because it is a single-exon gene,
and pangene puts \emph{NUDT4B} at the \emph{NUDT4} locus
because the protein sequence of \emph{NUDT4} is a strict substring of \emph{NUDT4B}.

We also compared the pangene annotation to the T2T-CHM13 annotation generated by Ensembl.
There are 41 differences out 18,676 genes present in both annotations.
These include four genes \emph{CSAG2}, \emph{CSAG3}, \emph{MAGEA6} and \emph{MAGEA3} that are close to each other.
Pangene gives better protein-to-genome alignment for the four genes.
We speculate that Ensembl may be annotating these genes based on synteny with GRCh38,
but synteny here might be broken due to an apparent inversion around the genes.
For another example, in the RCCX gene cluster~\citep{Carrozza:2021aa}, pangene annotates \emph{CYP21A2} ahead of \emph{TNXB}
but Ensembl puts \emph{CYP21A2} inside an intron of \emph{TNXB}, which seems to be an error.
Ensembl also misses \emph{C4A}, another gene in RCCX.
We believe pangene is doing better in these cases.
Most other genes among the 41 differences come from huge complex gene clusters.
We cannot tell what the correct annotation is.

We next investigated genes with presence/absence variants (PAVs).
Out of 16,996 multi-exon genes in the graph,
16,315 were on the autosomes of GRCh38.
3.4\% of them were absent from one or more haplotypes and 0.4\% ($n$=58) were only present in $\le$50\% of haplotypes.
Among the 58 genes, \emph{TRIM64}, for example, is close paralogous to \emph{TRIM64B}.
On most haplotypes, \emph{TRIM64B} was aligned better than \emph{TRIM64} and pangene annotated two \emph{TRIM64B} genes but no \emph{TRIM64}.
Therefore \emph{TRIM64} became a PAV.
When we excluded genes having paralogs of $\ge$95\% identity from the list, only 14 were left.
This list included three KIR genes and \emph{HLA-DRB4} that are known to be PAVs~\citep{Zhou2024.01.20.576452}.
We checked a few remaining genes in the list.
\emph{PRH1} and \emph{GSTM1} were in relatively simple regions and look like real PAVs.
Both of them have paralogs of identity below 95\%.
\emph{GSTT2} has a close paralog \emph{GSTT2B} that CD-HIT failed to identify -- it seemed an algorithm glitch.
The flanking region around \emph{NOTCH2NLR} looked complex and was poorly assembled in most haplotypes.
This may explain the low occurrence rate of this gene.
Overall, almost all human genes or their close paralogs are present in the majority of human haplotypes.
It is rare to see a protein-coding gene with sequences completely missing without paralogs as backups.

Our analysis is insensitive to thresholds $r_1$ and $r_2$ in Section~\ref{sec:adj}.
Reducing $r_1$ from 95\% to 90\% and $r_2$ from 98\% to 96\% only affects 9 out of 266 bibubbles.

\subsection{Incorporating 10 great ape haplotypes}

We also constructed a graph including ten haplotype assemblies of great apes, including chimpanzee, bonobo, gorilla
and two orangutan subspecies~\citep{Makova2023.11.30.569198}.
We filtered out single-exon genes and pruned edges supported by one contig.
This graph contained 239 generalized bibubbles, 131 of which were polymorphic among the 100 human samples.
The fewer polymorphic bibubbles in human (in comparison to 163 identified without great ape samples)
were caused by the merging of small bibubbles,
by the increased difficulty in resolving paralogs given more remote outgroups,
or by chromosome-scale inversions or translocations that disrupted the local bubble topology.
The last cause seemed to be the major contributor.
We will use the 17q21.31 inversion around \emph{MAPT} as an example.

The 17q21.31 inversion is polymorphic in the human population~\citep{Boettger:2012aa,Steinberg:2012aa}.
In the graph derived from the 100 human haplotypes, pangene identifid a generalized bibubble corresponding to this inversion.
The bibubble contained \emph{LRRC37A} and \emph{LRRC37A2} which have similar sequences.
The two genes have another paralog \emph{LRRC37A3} that is $\sim$18 Mb away.
All great apes only have one copy of the gene.
Intriguingly, gorilla haplotypes are similar to human haplotypes at 17q24.1
but chimpanzee and orangutan haplotypes involve genes at 17q21.31 (Fig.~\ref{fig:17inv}).
As a result, the pangene graph including great apes connected genes in 17q24.1 and 17q21.31
and broke the 17q21.31 bibubble in the human-only graph.
Involving outgroup species may hurt the finding of localized bibubbles.

\begin{table}[!tb]
\caption{Number of detected genes in bacterial pangenomes\label{tab:bac}}
\begin{tabular*}{\columnwidth}{@{\extracolsep\fill}lrrr@{\extracolsep\fill}}
\toprule
& Panaroo & pangene & PPanGGOLiN \\
\midrule
Mtb: \#total genes   & 4,207  & 4,216  & 4,744  \\
Mtb: \#core genes    & 3,769  & 3,774  & 3,660  \\
Ecoli: \#total genes & 14,342 & 13,902 & 14,291 \\
Ecoli: \#core genes  & 3,065  & 3,118  & 3,038  \\
\botrule
\end{tabular*}
\begin{tablenotes}
\item Panaroo, pangene and PPanGGOLiN were applied to two sets of gapless bacterial assemblies:
152 \emph{M. tuberculosis} (Mtb) strains and 50 \emph{E. coli} strains.
Mtb genes were annotated by Prokka and Ecoli genes were annotated by NCBI and downloaded from GenBank.
A ``core'' gene is a gene that is inferred to be present in $\ge$99\% of assemblies in each dataset.
Panaroo was invoked in the strict mode with paralogs merged (option ``{\tt
--clean-mode strict --merge\_paralogs}'').
\end{tablenotes}
\end{table}

\subsection{Analyzing 152 \textit{M. tuberculosis} strains}

We downloaded the \emph{M. tuberculosis} reference strain H37Rv and its gene annotation from RefSeq (AC:GCF\_000195955.2)
and obtained the complete long-read assemblies of 151 other strains~\citep{Peker:2021aa,Marin:2022aa,Hall:2023aa}.
Following the instruction of Panaroo~\citep{Tonkin-Hill:2020aa},
we ran Prodigal~\citep{Hyatt:2010aa} v2.6.3 on the reference strain to train the Prodigal model
and ran Prokka~\citep{Seemann:2014aa} v1.14.6 with the pretrained model to predict protein coding genes in all strains.
We used CD-HIT~\citep{Li:2006aa,Fu:2012aa} v4.8.1 with option ``{\tt -c 0.98}'' to cluster non-reference protein sequences,
which resulted in 6,744 clusters.
We mapped these proteins to each \emph{M. tuberculosis} genome using miniprot with option ``{\tt -S}'' to disable splicing.
We finally ran pangene with ``{\tt -p.001}''
to keep all genes regardless of their frequency in the pangenome.

Pangene constructed a graph consisting of 4,216 genes,
3,652 of which were present in all 152 genomes (Table~\ref{tab:bac}; Fig.~\ref{fig:bac-venn}).
To check if pangene captured the gene content in these strains,
we compared the pangene result to Panaroo v1.3.4.
We aligned the Panaroo proteins to the pangene proteins with MMseqs2~\citep{Steinegger:2017aa} v13.45111
and identified 46 Panaroo proteins do not hit to pangene proteins at the default E-value threshold of $10^{-3}$.
We mapped the 46 proteins to H37Rv with miniprot
and found 43 of them can be aligned
and 76\% of the aligned regions overlap with annotated CDS in RefSeq.
Manually investigating the overlaps revealed that most of the 43 proteins
were aligned to the opposite strand of some RefSeq genes or in different reading frames.
Identifying homology based on the genomic locations of input proteins,
pangene did not include them into the final graph.
Conversely, 40 pangene proteins do not hit to Panaroo proteins.

We additionally ran PPanGGOLiN~\citep{Gautreau:2020aa} v1.2.105 with the Prokka annotation as the input.
PPanGGOLiN collected 4,744 genes in the pangenome with 3,463 present in all.
277 genes did not hit to genes selected by pangene.
274 of these genes could be aligned to H37Rv by miniprot
and 90\% of the aligned regions overlap with annotated CDS in RefSeq.
We note that although overlapping genes in different open reading frames are rare in H37Rv (0.2\% of coding regions),
these genes may occur in other strains and may have functional implications~\citep{Vargas:2023aa,Snobre:2024aa}.
PPanGGOLiN's preference to count such genes in the pangenome is not necessarily a weakness.

\subsection{Analyzing 50 \textit{E. coli} strains}

\emph{M. tuberculosis} has low diversity with each strain similar to each other~\citep{Marin:2022aa}.
To understand how pangene performs given more diverse strains,
we collected 50 \emph{E. coli} genomes with complete assemblies~\citep{Shaw:2021aa}.

\begin{figure}[t!]
\centering
\includegraphics[width=.49\textwidth]{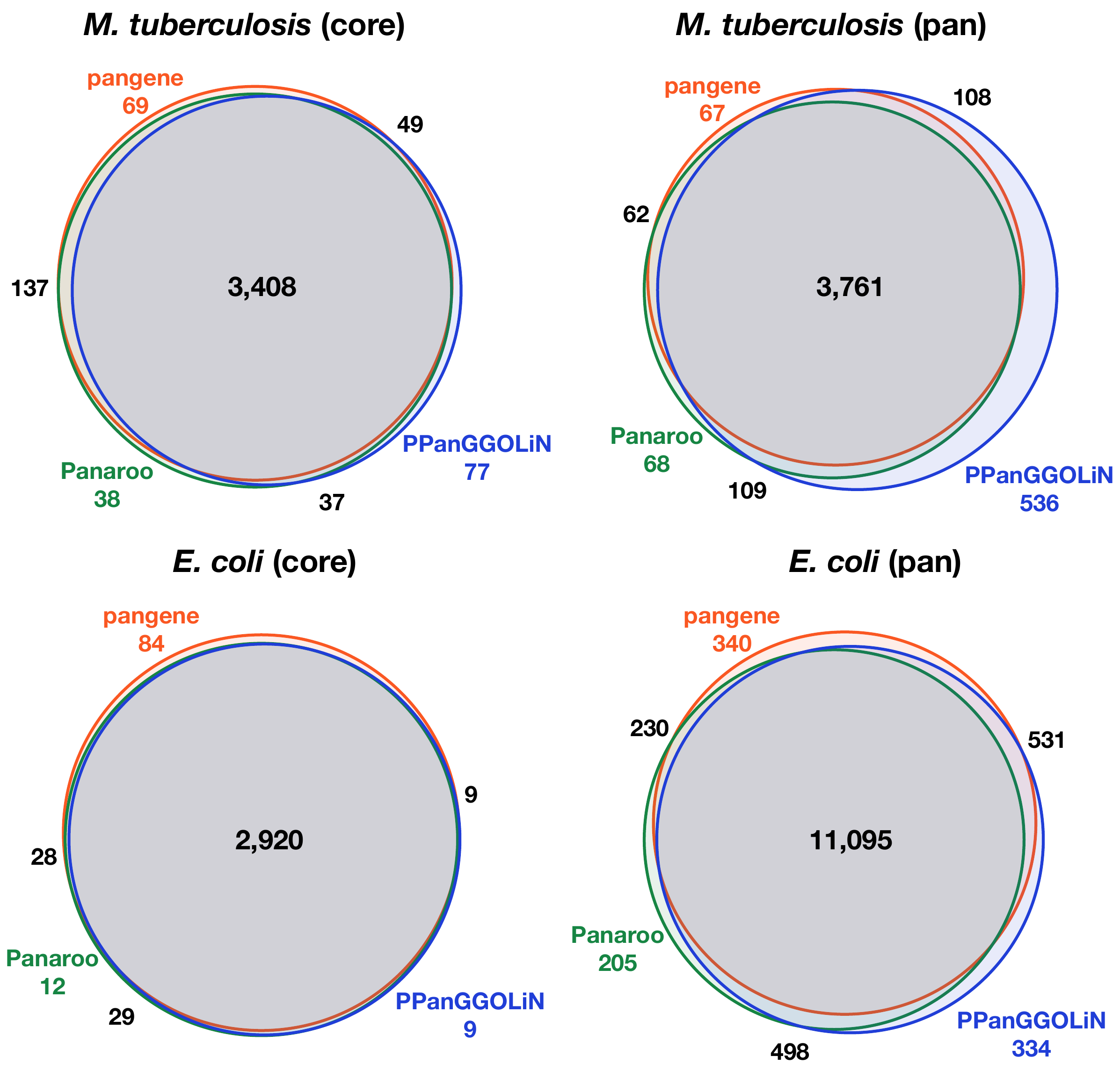}
\caption{Comparing three bacterial pangenome construction tools.
In each venn diagram, pangene, Panaroo and PPanGGOLiN proteins are merged and
clustered with MMseqs2 ({\tt easy-cluster -c 0.9 --min-seq-id 0.5}).
The diagram shows the number of clusters in each intersection.
MMseqs2 merges paralogous genes, so the number of clusters for each tool
is smaller than the number of genes in Table~\ref{tab:bac}.
Venn diagrams were generated with BioVenn~\citep{Hulsen:2008aa}.
}\label{fig:bac-venn}
\end{figure}

We followed the same procedure to run pangenome tools.
To get clean graph, pangene by default filters out genes that had $>$10 edges or connected $>$3 distant loci in the graph.
This filter only removed six genes in \emph{M. tuberculosis} dataset but
it filtered out 848 genes in \emph{E. coli}.
We added option ``{\tt -p.001 -g50 -r10}'' to retain more high-copy genes and genes connecting multiple loci.
Although pangene collected fewer genes (Table~\ref{tab:bac}),
only 72 Panaroo genes do not hit to genes collected by pangene
and the pangene pangenome size after clustering is not noticeably smaller than the Panaroo and the PPanGGOLiN pangenomes (Fig.~\ref{fig:bac-venn}).
The differences between the tools might be determined by subtle thresholds on how to resolve homology
and may not reflect the capability of each algorithm.

\section{Discussions}

Pangene constructs a graph to jointly represent the gene content of multiple high-quality genomes.
It provides approximate gene annotation and orthology assignment as a byproduct.
Given bacterial genomes, it plays a similar role to other bacterial pangenome tools
for identifying core and accessory genes~\citep{Tonkin-Hill:2023aa}.
Pangene is unique in that it also works for eukaryotic genomes
and captures localized gene-level variants.
Pangene complements whole-genome alignment based pangenome tools
and gives a more concise high-level view of gene content changes.

Constructing a pangene graph is challenging because the biologically correct graph is unknown.
We tend to think a ``correct'' graph should model the evolution history in principle.
However, evolution is complex in the presence of convergent/divergent evolution in the long time scale
and non-homologous recombinations and non-allelic gene conversions~\citep{Vollger:2023aa} in the short time scale.
We do not know the evolution history of most complex loci except several well-studied cases.
Furthermore, we take each gene as an atomic unit but biologically,
a gene may be subjected to structural changes such as losing/gaining exons or fusing with another gene.
These events are difficult to model even if we had the exact annotation in all input genomes.

The pangene graph construction algorithm is \emph{ad hoc}.
We have manually tuned various heuristics to correctly encode known gene-level variants in human,
but may have overlooked other cases that pangene may fail to represent.
Ideally, we would prefer to model graph construction to a global optimization problem.
We have not been able to find such a formulation that can reliably encode known variations.
This limitation is not specific to pangene; it is also applied to other pangenome construction tools.

As protein-to-genome alignment can identify orthology within mammals or even vertebrates~\citep{Li:2023ac},
we could in principle apply pangene to more distantly related species.
We did construct a gene graph from the human and mouse reference genomes.
The graph revealed synteny blocks visible from the human-mouse whole-genome alignment,
but the topology of the whole graph looked complex due to frequent large-scale rearrangements over long evolutionary distance.
Many generalized bibubbles identified from this graph were due to real orthologous alignments missed by miniprot.
We will need another way at higher level to investigate such a graph.
Another practical challenge in building a cross-species graph is to select the set of input proteins.
When constructing the great ape graph, we tried to merge human and great ape protein sequences.
In the resultant graph, we observed thousands of alignments in the intergenic or intronic regions of GRCh38
which degraded the overall quality of the graph.
The issue will become more prominent when we take diverse genomes without high-quality gene annotation as input.
Nevertheless, we note that cross-species whole-genome alignment is also challenging and is perhaps no better than pangene.
We still believe with improvements, pangene can construct a cross-species graph that helps to reveal evolution in longer term.

On the theoretical side, this article presented a rigorous definition of ``bubble'' in a bidirected graph
but it did not find an efficient algorithm to identify such generalized bibubbles.
While the current implementation in pangene works for gene graphs containing $\sim$20,000 genes,
it will be slow for a minigraph-cactus or PGGB graph that contains tens of millions of nodes.
How to efficiently identify generalized bibubbles remains an open and critical problem.
Furthermore, ``bubbles'' cannot represent rearrangements at the chromosome scale.
How to explore a multi-species gene graph or an alignment graph in general
may be another interesting research topic.

\section*{Acknowledgements}

We are grateful to Pouria Salehi and Luke O'Connor for the helpful discussions about the definition of variants in a sequence graph.

\section*{Author contributions}

H.L. conceived the project, implemented the algorithms, analyzed the data and drafted the manuscript.
M.M. and M.R.F. compiled the bacteria datasets and contributed to data validation and interpretation.

\section*{Conflict of interest}

None declared.

\section*{Funding}

This work is supported by National Institute of Health grant R01HG010040 and U01HG010961 (to H.L.),
T15LM007092 (to M.M.) and R01AI155765 (to M.R.F.),
and by Chan-Zuckerberg Initiative (to H.L.).

\section*{Data availability}

Pangene source code available at \url{https://github.com/lh3/pangene};
pre-built pangene graphs can be downloaded from \url{https://zenodo.org/records/8118576}
and visualized at \url{https://pangene.bioinweb.org}.

\bibliographystyle{apalike}
{\sffamily\small
\bibliography{pangene}}

\begin{thebibliography}{}

\bibitem[Boettger et~al., 2012]{Boettger:2012aa}
Boettger, L.~M., Handsaker, R.~E., Zody, M.~C., and McCarroll, S.~A. (2012).
\newblock Structural haplotypes and recent evolution of the human 17q21.31
  region.
\newblock {\em Nat Genet}, 44:881--5.

\bibitem[Brankovic et~al., 2016]{DBLP:journals/tcs/BrankovicIKMPV16}
Brankovic, L., Iliopoulos, C.~S., Kundu, R., et~al. (2016).
\newblock Linear-time superbubble identification algorithm for genome assembly.
\newblock {\em Theor. Comput. Sci.}, 609:374--383.

\bibitem[Carrozza et~al., 2021]{Carrozza:2021aa}
Carrozza, C., Foca, L., De~Paolis, E., and Concolino, P. (2021).
\newblock Genes and pseudogenes: Complexity of the {RCCX} locus and disease.
\newblock {\em Front Endocrinol (Lausanne)}, 12:709758.

\bibitem[Cheng et~al., 2021]{Cheng:2021aa}
Cheng, H., Concepcion, G.~T., Feng, X., et~al. (2021).
\newblock Haplotype-resolved de novo assembly using phased assembly graphs with
  hifiasm.
\newblock {\em Nat Methods}, 18:170--175.

\bibitem[Chin et~al., 2023]{Chin:2023aa}
Chin, C.-S., Behera, S., Khalak, A., et~al. (2023).
\newblock Multiscale analysis of pangenomes enables improved representation of
  genomic diversity for repetitive and clinically relevant genes.
\newblock {\em Nat Methods}, 20:1213--1221.

\bibitem[Dabbaghie et~al., 2022]{Dabbaghie:2022aa}
Dabbaghie, F., Ebler, J., and Marschall, T. (2022).
\newblock {BubbleGun}: enumerating bubbles and superbubbles in genome graphs.
\newblock {\em Bioinformatics}, 38:4217--4219.

\bibitem[Ding et~al., 2018]{Ding:2018aa}
Ding, W., Baumdicker, F., and Neher, R.~A. (2018).
\newblock panx: pan-genome analysis and exploration.
\newblock {\em Nucleic Acids Res}, 46:e5.

\bibitem[Fu et~al., 2012]{Fu:2012aa}
Fu, L. et~al. (2012).
\newblock {CD-HIT}: accelerated for clustering the next-generation sequencing
  data.
\newblock {\em Bioinformatics}, 28:3150--2.

\bibitem[Garrison et~al., 2023]{Garrison2023.04.05.535718}
Garrison, E., Guarracino, A., Heumos, S., et~al. (2023).
\newblock Building pangenome graphs.
\newblock {\em bioRxiv}.

\bibitem[G{\"a}rtner et~al., 2018]{Gartner:2018aa}
G{\"a}rtner, F., M{\"u}ller, L., and Stadler, P.~F. (2018).
\newblock Superbubbles revisited.
\newblock {\em Algorithms Mol Biol}, 13:16.

\bibitem[G{\"{a}}rtner and Stadler, 2019]{DBLP:journals/algorithms/GartnerS19}
G{\"{a}}rtner, F. and Stadler, P.~F. (2019).
\newblock Direct superbubble detection.
\newblock {\em Algorithms}, 12:81.

\bibitem[Gautreau et~al., 2020]{Gautreau:2020aa}
Gautreau, G. et~al. (2020).
\newblock {PPanGGOLiN}: Depicting microbial diversity via a partitioned
  pangenome graph.
\newblock {\em PLoS Comput Biol}, 16:e1007732.

\bibitem[Hall et~al., 2023]{Hall:2023aa}
Hall, M.~B., Rabodoarivelo, M.~S., Koch, A., et~al. (2023).
\newblock Evaluation of {Nanopore} sequencing for {Mycobacterium} tuberculosis
  drug susceptibility testing and outbreak investigation: a genomic analysis.
\newblock {\em Lancet Microbe}, 4:e84--e92.

\bibitem[Handsaker et~al., 2015]{Handsaker:2015ur}
Handsaker, R.~E. et~al. (2015).
\newblock Large multiallelic copy number variations in humans.
\newblock {\em Nat Genet}, 47:296--303.

\bibitem[He et~al., 2023]{He:2023aa}
He, Y., Chu, Y., Guo, S., et~al. (2023).
\newblock {T2T-YAO}: A telomere-to-telomere assembled diploid reference genome
  for han chinese.
\newblock {\em Genomics Proteomics Bioinformatics}.

\bibitem[Hickey et~al., 2023]{Hickey:2023aa}
Hickey, G., Monlong, J., Ebler, J., et~al. (2023).
\newblock Pangenome graph construction from genome alignments with
  minigraph-cactus.
\newblock {\em Nat Biotechnol}.

\bibitem[Hulsen et~al., 2008]{Hulsen:2008aa}
Hulsen, T., de~Vlieg, J., and Alkema, W. (2008).
\newblock {BioVenn} - a web application for the comparison and visualization of
  biological lists using area-proportional venn diagrams.
\newblock {\em BMC Genomics}, 9:488.

\bibitem[Hyatt et~al., 2010]{Hyatt:2010aa}
Hyatt, D. et~al. (2010).
\newblock Prodigal: prokaryotic gene recognition and translation initiation
  site identification.
\newblock {\em BMC Bioinformatics}, 11:119.

\bibitem[Johnson et~al., 1994]{DBLP:conf/pldi/JohnsonPP94}
Johnson, R., Pearson, D., and Pingali, K. (1994).
\newblock The program structure tree: Computing control regions in linear time.
\newblock In Sarkar, V., Ryder, B.~G., and Soffa, M.~L., editors, {\em
  Proceedings of the {ACM} SIGPLAN'94 Conference on Programming Language Design
  and Implementation (PLDI), Orlando, Florida, USA, June 20-24, 1994}, pages
  171--185. {ACM}.

\bibitem[Ju et~al., 2016]{Ju:2016aa}
Ju, X.-C., Hou, Q.-Q., Sheng, A.-L., et~al. (2016).
\newblock The hominoid-specific gene {TBC1D3} promotes generation of basal
  neural progenitors and induces cortical folding in mice.
\newblock {\em Elife}, 5.

\bibitem[Li, 2023]{Li:2023ac}
Li, H. (2023).
\newblock Protein-to-genome alignment with miniprot.
\newblock {\em Bioinformatics}, 39:btad014.

\bibitem[Li et~al., 2020]{Li:2020aa}
Li, H. et~al. (2020).
\newblock The design and construction of reference pangenome graphs with
  minigraph.
\newblock {\em Genome Biol}, 21:265.

\bibitem[Li and Godzik, 2006]{Li:2006aa}
Li, W. and Godzik, A. (2006).
\newblock {CD-HIT}: a fast program for clustering and comparing large sets of
  protein or nucleotide sequences.
\newblock {\em Bioinformatics}, 22:1658--9.

\bibitem[Liao et~al., 2023]{Liao:2023aa}
Liao, W.-W., Asri, M., Ebler, J., et~al. (2023).
\newblock A draft human pangenome reference.
\newblock {\em Nature}, 617:312--324.

\bibitem[Makova et~al., 2023]{Makova2023.11.30.569198}
Makova, K.~D., Pickett, B.~D., Harris, R.~S., et~al. (2023).
\newblock The complete sequence and comparative analysis of ape sex
  chromosomes.
\newblock {\em bioRxiv}.

\bibitem[Marin et~al., 2022]{Marin:2022aa}
Marin, M. et~al. (2022).
\newblock Benchmarking the empirical accuracy of short-read sequencing across
  the m. tuberculosis genome.
\newblock {\em Bioinformatics}, 38:1781--1787.

\bibitem[Mercuri et~al., 2022]{Mercuri:2022aa}
Mercuri, E., Sumner, C.~J., Muntoni, F., Darras, B.~T., and Finkel, R.~S.
  (2022).
\newblock Spinal muscular atrophy.
\newblock {\em Nat Rev Dis Primers}, 8:52.

\bibitem[Nurk et~al., 2022]{Nurk:2022up}
Nurk, S. et~al. (2022).
\newblock The complete sequence of a human genome.
\newblock {\em Science}, 376:44--53.

\bibitem[Nurk et~al., 2020]{Nurk:2020we}
Nurk, S., Walenz, B.~P., Rhie, A., et~al. (2020).
\newblock {HiCanu}: accurate assembly of segmental duplications, satellites,
  and allelic variants from high-fidelity long reads.
\newblock {\em Genome Res}, 30:1291--1305.

\bibitem[Onodera et~al., 2013]{DBLP:conf/wabi/OnoderaSS13}
Onodera, T., Sadakane, K., and Shibuya, T. (2013).
\newblock Detecting superbubbles in assembly graphs.
\newblock In {\em {WABI}}, pages 338--348.

\bibitem[Page et~al., 2015]{Page:2015aa}
Page, A.~J., Cummins, C.~A., Hunt, M., et~al. (2015).
\newblock Roary: rapid large-scale prokaryote pan genome analysis.
\newblock {\em Bioinformatics}, 31:3691--3.

\bibitem[Paten et~al., 2018]{Paten:2018aa}
Paten, B., Eizenga, J.~M., Rosen, Y.~M., et~al. (2018).
\newblock Superbubbles, ultrabubbles, and cacti.
\newblock {\em J Comput Biol}, 25:649--663.

\bibitem[Peker et~al., 2021]{Peker:2021aa}
Peker, N., Schuele, L., Kok, N., et~al. (2021).
\newblock Evaluation of whole-genome sequence data analysis approaches for
  short- and long-read sequencing of mycobacterium tuberculosis.
\newblock {\em Microb Genom}, 7.

\bibitem[Rautiainen et~al., 2023]{Rautiainen:2023aa}
Rautiainen, M., Nurk, S., Walenz, B.~P., et~al. (2023).
\newblock Telomere-to-telomere assembly of diploid chromosomes with {Verkko}.
\newblock {\em Nat Biotechnol}.

\bibitem[Schneider et~al., 2017]{Schneider:2017aa}
Schneider, V.~A. et~al. (2017).
\newblock Evaluation of {GRCh38} and de novo haploid genome assemblies
  demonstrates the enduring quality of the reference assembly.
\newblock {\em Genome Res}, 27:849--864.

\bibitem[Seemann, 2014]{Seemann:2014aa}
Seemann, T. (2014).
\newblock Prokka: rapid prokaryotic genome annotation.
\newblock {\em Bioinformatics}, 30:2068--9.

\bibitem[Shaw et~al., 2021]{Shaw:2021aa}
Shaw, L.~P. et~al. (2021).
\newblock Niche and local geography shape the pangenome of wastewater- and
  livestock-associated {Enterobacteriaceae}.
\newblock {\em Sci Adv}, 7.

\bibitem[Snobre et~al., 2024]{Snobre:2024aa}
Snobre, J., Meehan, C.~J., Mulders, W., et~al. (2024).
\newblock Frameshift mutations in {Rv0678} preserve bedaquiline susceptibility
  in {Mycobacterium} tuberculosis by maintaining protein integrity.
\newblock {\em SSRN}.

\bibitem[Steinberg et~al., 2012]{Steinberg:2012aa}
Steinberg, K.~M., Antonacci, F., Sudmant, P.~H., et~al. (2012).
\newblock Structural diversity and african origin of the 17q21.31 inversion
  polymorphism.
\newblock {\em Nat Genet}, 44:872--80.

\bibitem[Steinegger and S{\"o}ding, 2017]{Steinegger:2017aa}
Steinegger, M. and S{\"o}ding, J. (2017).
\newblock Mmseqs2 enables sensitive protein sequence searching for the analysis
  of massive data sets.
\newblock {\em Nat Biotechnol}, 35:1026--1028.

\bibitem[Sudmant et~al., 2010]{Sudmant:2010aa}
Sudmant, P.~H. et~al. (2010).
\newblock Diversity of human copy number variation and multicopy genes.
\newblock {\em Science}, 330:641--6.

\bibitem[Sung et~al., 2015]{Sung:2015aa}
Sung, W.-K., Sadakane, K., Shibuya, T., et~al. (2015).
\newblock An o(m log m)-time algorithm for detecting superbubbles.
\newblock {\em IEEE/ACM Trans Comput Biol Bioinform}, 12:770--7.

\bibitem[Taylor et~al., 2020]{Taylor:2020aa}
Taylor, C., Crosby, I., Yip, V., et~al. (2020).
\newblock A review of the important role of {CYP2D6} in pharmacogenomics.
\newblock {\em Genes (Basel)}, 11.

\bibitem[Tonkin-Hill et~al., 2023]{Tonkin-Hill:2023aa}
Tonkin-Hill, G., Corander, J., and Parkhill, J. (2023).
\newblock Challenges in prokaryote pangenomics.
\newblock {\em Microb Genom}, 9.

\bibitem[Tonkin-Hill et~al., 2020]{Tonkin-Hill:2020aa}
Tonkin-Hill, G. et~al. (2020).
\newblock Producing polished prokaryotic pangenomes with the panaroo pipeline.
\newblock {\em Genome Biol}, 21:180.

\bibitem[Vargas et~al., 2023]{Vargas:2023aa}
Vargas, Jr, R., Luna, M.~J., Freschi, L., et~al. (2023).
\newblock Phase variation as a major mechanism of adaptation in {Mycobacterium}
  tuberculosis complex.
\newblock {\em Proc Natl Acad Sci U S A}, 120:e2301394120.

\bibitem[Vollger et~al., 2023]{Vollger:2023aa}
Vollger, M.~R., Dishuck, P.~C., Harvey, W.~T., et~al. (2023).
\newblock Increased mutation and gene conversion within human segmental
  duplications.
\newblock {\em Nature}, 617:325--334.

\bibitem[Wenger et~al., 2019]{Wenger_2019}
Wenger, A.~M., Peluso, P., Rowell, W.~J., et~al. (2019).
\newblock Accurate circular consensus long-read sequencing improves variant
  detection and assembly of a human genome.
\newblock {\em Nat Biotechnol}, 37:1155--1162.

\bibitem[Wick et~al., 2015]{Wick:2015qf}
Wick, R.~R., Schultz, M.~B., Zobel, J., and Holt, K.~E. (2015).
\newblock Bandage: interactive visualization of de novo genome assemblies.
\newblock {\em Bioinformatics}, 31:3350--2.

\bibitem[Yang et~al., 2023]{Yang:2023aa}
Yang, C., Zhou, Y., Song, Y., et~al. (2023).
\newblock The complete and fully-phased diploid genome of a male han chinese.
\newblock {\em Cell Res}, 33:745--761.

\bibitem[Zerbino and Birney, 2008]{Zerbino:2008uq}
Zerbino, D.~R. and Birney, E. (2008).
\newblock Velvet: algorithms for de novo short read assembly using de {Bruijn}
  graphs.
\newblock {\em Genome Res}, 18:821--9.

\bibitem[Zhou et~al., 2024]{Zhou2024.01.20.576452}
Zhou, Y., Song, L., and Li, H. (2024).
\newblock Full resolution {HLA} and {KIR} genes annotation for human genome
  assemblies.
\newblock {\em bioRxiv}.

\bibitem[Zhou et~al., 2020]{Zhou:2020aa}
Zhou, Z., Charlesworth, J., and Achtman, M. (2020).
\newblock Accurate reconstruction of bacterial pan- and core genomes with
  {PEPPAN}.
\newblock {\em Genome Res}, 30:1667--1679.

\end{thebibliography}

\end{document}